\documentclass[12pt]{article}
\usepackage{amsfonts,amsmath,graphicx}

\def\ie{{\it i.e.}}
\def\Rop{{\mathbb{R}}}
\def\Cop{{\mathbb{C}}}
\def\Zop{{\mathbb{Z}}}
\def\Nop{{\mathbb{N}}}
\def\bH{\pmb{\cal H}}
\def\H{{\cal H}}
\def\bz{\bar{z}}
\newcommand\be{\begin{equation}}
\newcommand\ee{\end{equation}}

\begin{document}

\pagestyle{myheadings}
\markright{2D conformal field theory and vertex operator algebras}

\parindent 0mm
\parskip 6pt


\title{2D conformal field theory and vertex operator algebras}

\author{Matthias R. Gaberdiel\\
Institut f\"ur Theoretische Physik \\
ETH Z\"urich, ETH-H\"onggerberg \\
8093 Z\"urich, Switzerland \\
E-mail: gaberdiel@itp.phys.ethz.ch}

\date{2 September 2005}

\maketitle


\section*{Introduction}

For the last twenty years or so two-dimensional conformal field
theories have played an important role in different areas of modern
theoretical physics. One of the main applications of conformal field
theory has been in string theory (see the articles on string and
superstring theory), where the excitations of the string
are described, from the point of view of the world-sheet, by a 
two-dimensional conformal field theory. Conformal field theories have
also been studied in the context of statistical physics since the 
critical points of second order phase transition are typically
described by a  conformal field theory. Finally, conformal field
theories are interesting solvable toy models of genuinely interacting
quantum field theories.  

{}From an abstract point of view, conformal field theories are
(Euclidean) quantum field theories that are characterized by the
property that their symmetry group contains, in addition to the
Euclidean symmetries, local conformal transformations, \ie\
transformations that preserve angles but not necessarily lengths. The
local conformal symmetry is of special importance in two dimensions
since the corresponding symmetry algebra is infinite-dimensional in
this case. As a consequence, two-dimensional conformal field theories
have an infinite number of conserved quantities, and are 
essentially solvable by symmetry considerations alone. The
mathematical formulation of these symmetries has led to the concept of
a vertex operator algebra that has become a new branch of
mathematics in its own right. In particular, it has played a major
role in the explanation of `monstrous moonshine' for which Richard
Borcherds received the fields medal in 1998.

\section*{The conformal symmetry group}

The conformal symmetry group of the $n$-dimensional Euclidean space
$\Rop^n$ consists of the (locally defined) transformations that
preserve angles but not necessarily lengths. 
The transformations that preserve angles as well as lengths
are the well-known translations and rotations. The conformal group 
contains (in any dimension) in addition the {\it dilatations} or scale
transformations 
\be
x^\mu\mapsto \tilde x^\mu = \lambda x^\mu \,,
\ee
where $\lambda\in\Rop$ and $x^\mu \in \Rop^n$, as well as the 
so-called {\it special conformal transformations}, 
\be
x^\mu\mapsto \tilde x^\mu = 
\frac{x^{\mu} + {\bf x}^2\, a^{\mu}}
{1 + 2 ({\bf x}\cdot{\bf a}) + {\bf x}^2\, {\bf a}^2} \,,
\ee
where $a^\mu\in\Rop^n$ and ${\bf x}^2 = x^\mu x_\mu$. (Note that this 
last transformation is only defined for 
$x^\mu\neq - a^\mu/{\bf a}^2$.) 

If the dimension $n$ of the space $\Rop^n$  is bigger than two, one
can show that the full conformal group is generated by these
transformations. For $n=2$, however, the group of (locally defined)
conformal transformations is much larger. To see this it is
convenient to introduce complex coordinates for $(x,y)\in \Rop^2$ by
defining $z=x+iy$ and $\bar{z} = x-iy$. Then any (locally) 
{\it analytic} function $f(z)$ defines a conformal transformation by
$z\mapsto f(z)$, since analytic maps preserve angles. (Incidentally,
the same also applies to $z\mapsto \overline{f(z)}$, but this would
reverse the orientation.)  Clearly, the group of such transformations
is infinite dimensional; this is a special feature of two dimensions.

In this complex notation, the transformations that are generated by
translations, rotations, dilatations and special conformal
transformations simply generate the M\"obius group of automorphisms of
the Riemann sphere
\be\label{e1}
z \mapsto f(z) = \frac{az + b}{cz+d} \,,
\ee
where $a,b,c,d$ are complex constants with $ad-bc\ne 0$; since
rescaling $a,b,c,d$ by a common complex number does not modify
(\ref{e1}), the M\"obius group is isomorphic to SL$(2,\Cop)/\Zop_2$. 
In addition to these transformations (that are globally defined on the
Riemann sphere), we have an infinite set of infinitesimal
transformations generated by 
$L_n: z\mapsto z+\epsilon z^{n+1}$ for all $n\in\Zop$. The generators
$L_{\pm 1}$ and $L_0$ generate the subgroup of M\"obius
transformations, and their commutation relations are simply 
\be\label{witt}
{}[L_m,L_n] = (m-n)\, L_{m+n} \,.
\ee
In fact, (\ref{witt}) describes also the commutation relations of all 
generators $L_n$ with $n\in\Zop$: this is the Lie algebra of
(locally defined) two-dimensional conformal transformations --- it is
called the {\it Witt algebra}.

\section*{The general structure of conformal field theory} 

A two-dimensional conformal field theory is determined (like any other
field theory) by its space of states and the collection of its
correlation functions (vacuum expectation values). The space of states
is a vector space $\bH$ (that is in many interesting examples a
Hilbert space), and the correlation functions are defined for
collections of vectors in 
some dense subspace of $\bH$. These correlation functions are defined
on a two-dimensional (Euclidean) space. We shall mainly be interested 
in the case where the underlying two-dimensional space is a closed
compact surface; the other important case concerning surfaces with
boundaries (whose analysis was pioneered by Cardy) will
be reviewed elsewhere (see the article on `Boundary conformal field
theory'). The closed surfaces are classified 
(topologically) by their genus $g$ which counts the number of handles;
the simplest such surface which we shall mainly consider is the sphere
with $g=0$, the surface with $g=1$ is the torus, {\it etc.}  

One of the special features of conformal field theory is the fact that
the theory is naturally defined on a {\em Riemann surface} (or 
{\em complex curve}), {\it i.e.} on a surface that possesses suitable
complex coordinates. In the case of the sphere, the complex
coordinates can be taken to be those of the complex plane that cover
the sphere except for the point at infinity; complex coordinates
around infinity are defined by means of the coordinate function
$\gamma(z)=1/z$ that maps a neighborhood of infinity to a
neighborhood of zero. With this choice of complex coordinates, the
sphere is usually referred to as the {\em Riemann sphere}, and this
choice of complex coordinates is up to M\"obius transformations
unique. The correlation functions of a conformal field theory that is
defined on the sphere are thus of the form    
\be
\label{loccorr}
\langle 0|V(\psi_1;z_1,\bar{z}_1) \cdots V(\psi_n;z_n,\bar{z}_n) 
|0 \rangle \,,
\ee
where $V(\psi,z,\bar{z})$ is the field that is associated to the state
$\psi$, and $z_i$ and $\bar{z}_i$ are complex conjugates of one
another. Here $|0\rangle$ denotes the SL$(2,\Cop)/\Zop_2$-invariant
vacuum. The usual locality assumption of a two-dimensional (bo\-so\-nic)
Euclidean quantum field theory implies that these correlation
functions are independent of the order in which the fields appear in 
(\ref{loccorr}).  

It is conventional to think of $z=0$ as describing `past infinity',
and $z=\infty$ as `future infinity'; this defines a time direction in
the Euclidean field theory and thus a quantization scheme (radial
quantization). Furthermore we identify the space of states with the
space of `incoming' states; thus the state $\psi$ is simply
\be\label{state}
\psi=V(\psi;0,0) |0\rangle\,.
\ee

We can think of $z_i$ and $\bar{z}_i$ in (\ref{loccorr}) as
independent variables, \ie\ we may relax the constraint that
$\bar{z}_i$ is the complex conjugate of $z_i$. Then we have two
commuting actions of the conformal group on these correlations
functions: the infinitesimal action on the $z_i$ variables is
described (as before) by the $L_n$ generators, while the generators
for the action on the $\bar{z}_i$ variables are $\bar{L}_n$. In a
conformal field theory, the space of states $\bH$ thus carries two
commuting actions of the Witt algebra. The generator $L_0+\bar{L}_0$
can be identified with the time-translation operator, and thus
describes the energy operator. The space of states of the physical
theory should have a bounded energy spectrum, and it is thus natural
to assume that the spectrum of both  $L_0$ and $\bar{L}_0$ is bounded
from below; representations with this property are usually called {\it
positive energy representations}. It is relatively easy to see that
the Witt algebra does not have any unitary positive energy
representations except for the trivial representation. However, as is
common in many instances in quantum theory, it possesses many
interesting projective representations. These projective
representations are conventional representations of the central
extension of the Witt algebra 
\be\label{vira}
{}[L_m,L_n] = (m-n)\, L_{m+n} + \frac{c}{12}\, m \, (m^2-1) \,
\delta_{m,-n} \,,
\ee
which is the famous {\it Virasoro algebra}. Here $c$ is a central
element that commutes with all $L_m$; it is called the {\it central
charge} (or conformal anomaly).

Given the actions of the two Virasoro algebras (that are generated by
$L_n$ and $\bar{L}_n$), one can decompose the space of states $\bH$
into irreducible representations as 
\be\label{decomp}
\bH = \bigoplus_{ij} M_{ij}\,  \H_i \otimes \bar{\H}_j \,,
\ee
where $\H_i$ ($\bar{\H}_j$) denotes the irreducible representations of
the algebra of the $L_n$ ($\bar{L}_n$), and $M_{ij}\in\Nop_0$ describe
the multiplicities with which these combinations of representations
occur. [We are assuming here that the space of states is completely
reducible with respect to the action of the two Virasoro algebras;
examples where this is not the case are the so-called logarithmic
conformal field theories.] The positive energy representations of the
Virasoro algebra are 
characterized by the value of the central charge, as well as the
lowest eigenvalue of $L_0$; the state $\psi$ whose $L_0$ eigenvalue is
smallest is called the {\it highest weight state}, and its eigenvalue
$L_0 \psi = h \psi$ is the {\it conformal weight}. The conformal
weight determines the conformal transformation properties of $\psi$:
under the conformal transformation $z\mapsto f(z)$, 
$\bar{z} \mapsto \bar{f}(\bar{z})$, we have 
\be\label{primary}
V(\psi;z,\bar{z})  \mapsto 
\left(f'(z)\right)^h \, 
\left(\bar{f}'(\bar{z})\right)^{\bar{h}} \, 
V(\psi;f(z),\bar{f}(\bar{z})) \,,
\ee
where $L_0\psi = h \psi$ and $\bar{L}_0 \psi = \bar{h}\psi$. The
corresponding field $V(\psi;z,\bar{z})$ is then called a {\it primary}
field; if (\ref{primary}) only holds for the M\"obius transformations
(\ref{e1}), the field is called {\it quasiprimary}.  

Since $L_m$ with $m>0$ lowers the conformal weight of a state (see
(\ref{vira})), the highest weight state $\psi$ is necessarily
annihilated by all $L_m$ (and $\bar{L}_m$) with $m>0$. However, in
general the $L_m$ (and $\bar{L}_m$) with $m<0$ do not annihilate
$\psi$; they generate the {\it descendants} of $\psi$ that lie in the
same representation. Their conformal transformation property is more
complicated, but can be deduced from that of the primary state
(\ref{primary}), as well as the commutation relations of the Virasoro
algebra.

The M\"obius symmetry (whose generators annihilate the vacuum)
determines the 1-, 2- and 3-point functions of quasiprimary fields up
to numerical constants: the 1-point function vanishes, unless
$h=\bar{h}=0$, in which case 
$\langle 0| V(\psi;z,\bar{z})|0\rangle =C$, independent of $z$ and
$\bar{z}$. The 2-point function of $\psi_1$ and $\psi_2$ vanishes
unless $h_1=h_2$ and $\bar{h}_1=\bar{h}_2$; if the conformal weights
agree, it takes the form
\be
\langle 0 | V(\psi_1;z_1,\bar z_1)\,  V(\psi_2;z_2,\bar z_2) | 0 \rangle 
= C (z_1-z_2)^{-2 h}\, (\bar{z}_1 - \bar{z}_2)^{-2\bar{h}}  \,.
\ee
Finally, the structure of the 3-point function of three quasiprimary
fields $\psi_1$, $\psi_2$ and $\psi_3$ is 
\begin{eqnarray}
&& \qquad \langle 0 | V(\psi_1;z_1,\bar z_1)\,  V(\psi_2;z_2,\bar z_2) \,
V(\psi_3;z_3,\bar z_3) | 0 \rangle \nonumber \\
&& \qquad \qquad \qquad \qquad
= C \prod_{i<j} (z_i-z_j)^{(h_k - h_i - h_j)} \, 
(\bar{z}_i-\bar{z}_j)^{(\bar{h}_k - \bar{h}_i - \bar{h}_j)} \,,
\end{eqnarray}
where for each pair $i<j$, $k$ labels the third field, \ie\ 
$k\ne i$ and $k\ne j$. The M\"obius symmetry also restricts the higher
correlation function of quasiprimary fields: the 4-point function is
determined up to an (undetermined) function of the M\"obius invariant
cross-ratio, and similar statements also hold for $n$-point functions
with $n\geq 5$. The full Virasoro symmetry must then be used to
restrict these functions further; however, since the generators $L_n$
with $n\leq -2$ do not annihilate the vacuum $|0\rangle$, the Virasoro
symmetry leads to {\it Ward identities} that cannot be easily
evaluated in general. (In typical examples, these Ward identities
give rise to differential equations that must be obeyed by the
correlation functions.)

\section*{Chiral fields and vertex operator algebras}

The decomposition (\ref{decomp}) contains usually a special class of
states that transform as the vacuum state with respect to $\bar{L}_m$;
these states are the so-called {\it chiral} states. (Similarly, the
states that transform as the vacuum state with respect to $L_m$ are
the anti-chiral states.) Given the transformation properties described
above, it is not difficult to see that the corresponding chiral fields
$V(\psi;z,\bar{z})$ only depend on $z$ in any correlation function, 
\ie\ $V(\psi;z,\bar{z})\equiv V(\psi,z)$. (Similarly, the anti-chiral
fields only depend on $\bar{z}$.) The chiral fields always contain the
field corresponding to the state $L_{-2} |0\rangle$, that describes a
specific component of the stress-energy tensor. 

In conformal field theory the product of two fields can be
expressed again in terms of the fields of the theory. The conformal
symmetry restricts the structure of this operator product
expansion:
\begin{eqnarray}
\label{localope}
&& V(\psi_1;z_1,\bar{z}_1) V(\psi_2;z_2,\bar{z}_2)  \\
& & \quad = \sum_{i} (z_1-z_2)^{\Delta_i} \,
(\bar{z}_1-\bar{z}_2)^{\bar\Delta_i}  \,
\sum_{r,s\geq 0} V(\phi_{r,s}^{i};z_2,\bar{z}_2) \, 
(z_1-z_2)^r \, (\bar{z}_1-\bar{z}_2)^s\,,\nonumber
\end{eqnarray}
where $\Delta_i$ and $\bar\Delta_i$ are real numbers, and
$r,s\in\Nop_0$. (Here $i$ labels the conformal representations that
appear in the operator product expansion, while $r$ and $s$ label the
different descendants.) The actual form of this expansion (in
particular, which representations appear) can be read off from the
correlation functions of the theory since the identity
(\ref{localope}) has to hold in {\it all} correlation functions.   

Given that the chiral fields only depend on $z$ in all correlation
functions, it is then clear that the operator product expansion of two
chiral fields contains again only chiral fields. Thus the subspace of
chiral fields closes under the operator product expansion, and
therefore defines a consistent (sub)theory by itself. This subtheory
is sometimes referred to as a meromorphic conformal field theory
(Goddard 1989). (Obviously, the same also applies to the subtheory of 
anti-chiral fields.) The operator product expansion defines a 
product on the space of meromorphic fields. This product involves the
complex parameters $z_i$ in a non-trivial way, and therefore does not
directly define an algebra structure; it is however very similar to an
algebra, and is therefore usually called a {\it vertex operator
algebra} in the mathematical literature. The formal definition
involves formal power series calculus and is quite complicated;
details can be found in (Frenkel-Lepowski-Meurman 1988). 
\smallskip

By virtue of its definition as an identity that holds in arbitrary
correlation functions, the operator product expansion is 
{\it associative}, \ie\  
\begin{eqnarray}\label{associative}
& & \Bigl( V(\psi_1;z_1,\bz_1) V(\psi_2;z_2,\bz_2) \Bigr) 
V(\psi_3;z_3,\bz_3) \\
& & \qquad \qquad \qquad \qquad \qquad 
= V(\psi_1;z_1,\bz_1) \Bigl(V(\psi_2;z_2,\bz_2)
V(\psi_3;z_3,\bz_3) \Bigr) \,, \nonumber 
\end{eqnarray}
where the brackets indicate which operator product expansion is
evaluated first. If we consider the case where both $\psi_1$ and
$\psi_2$ are meromorphic fields, then the associativity of the
operator product expansion implies that the states in $\bH$ form a
{\it representation} of the vertex operator algebra. The same also
holds for the vertex operator algebra associated to the
anti-chiral fields. Thus the meromorphic fields encode in a sense the
symmetries of the underlying theory: this symmetry always contains the
conformal symmetry (since $L_{-2}|0\rangle$ is always a chiral field,
and $\bar{L}_{-2}|0\rangle$ always an anti-chiral field). In general,
however, the symmetry may be larger. In order to take full advantage of
this symmetry, it is then useful to decompose the full space of states
$\bH$ not just with respect to the two Virasoro algebras, but rather
with respect to the two vertex operator algebras; the structure is
again the same as in (\ref{decomp}), where however now each $\H_i$ and
$\bar\H_j$ is an irreducible representation of the chiral and
anti-chiral vertex operator algebra, respectively.

\section*{Rational theories and Zhu's algebra}

Of particular interest are the rational conformal field theories that
are characterized by the property that the corresponding vertex
operator algebras only possess {\it finitely} many irreducible 
representations. (The name `rational' stems from the fact that the
conformal weights and the central charge of these theories are
rational numbers.)  The simplest example of such rational theories are
the so-called {\it minimal models}, for which the vertex operator
algebra describes just the conformal symmetry: these models exist for
a certain discrete set of central charges $c<1$ and were first studied
by Belavin, Polyakov and Zamolodchikov in 1984. (This paper is
contained in the reprint volume Goddard-Olive 1988.) It was this
seminal paper  that started much of the modern developments in
conformal field theory. Another important class of examples are the
Wess-Zumino-Witten (WZW) models that describe the world-sheet
theory of strings moving on a compact Lie group. The
relevant vertex operator algebra is then generated by the loop group
symmetries. There is some evidence that all rational conformal field
theories can be obtained from the WZW models by means of two 
standard constructions, namely by considering cosets and taking
orbifolds; thus rational conformal field theory seems
to have something of the flavor of (reductive) Lie theory. 

Rational theories may be characterized in terms of Zhu's algebra that
can be defined as follows. The chiral fields $V(\psi,z)$ that only
depend on $z$ must by themselves define local operators; they can
therefore be expanded in a Laurent expansion as 
\be
V(\psi,z) = \sum_{n\in\Zop} V_n(\psi)\, z^{-n-h} \,,
\ee
where $h$ is the conformal weight of the state $\psi$. For example,
for the case of the holomorphic component of the stress energy tensor
one finds 
\be
T(z) = \sum_{n\in\Zop} L_n z^{-n-2} \,,
\ee
where the $L_n$ are the Virasoro generators. By the state/field
correspondence (\ref{state}), it then follows that 
\be\label{e2}
V_n(\psi) |0\rangle = 0 \qquad \hbox{for $n > -h$} \,,
\ee
and that
\be\label{e3}
V_{-h}(\psi) |0\rangle = \psi \,.
\ee 
(For the example of the above component of the stress energy tensor, 
(\ref{e2}) implies that 
$L_{-1}|0\rangle = L_0|0\rangle  = L_{n}|0\rangle =0$ for $n\geq 0$ ---
thus the vacuum is in particular SL$(2,\Cop)/\Zop_2$
invariant. Furthermore, (\ref{e3}) shows that $L_{-2} |0\rangle$ is
the state corresponding to this component of the stress energy
tensor.) We denote by $\H_0$ the space of states that can be generated
by the action of the modes $V_n(\psi)$ from the vacuum $|0\rangle$. On
$\H_0$ we consider the subspace ${\cal O}(\H_0)$ that is spanned by
the states of the form
\be
V^{(N)}(\psi) \chi \,, \qquad N>0 \,,
\ee
where $V^{(N)}(\psi)$ is defined by 
\be
V^{(N)}(\psi) = \sum_{n=0}^{h} 
\left( \begin{array}{c} h\\ n \end{array} \right)
V_{-n-N}(\psi) \,,
\ee
and $h$ is the conformal weight of $\psi$. Zhu's algebra is
then the quotient space
\be
A = \H_0 / {\cal O}(\H_0) \,.
\ee
It actually forms an associative algebra, where the algebra structure
is defined by 
\be
\psi \star \chi = V^{(0)}(\psi)\, \chi \,.
\ee
This algebra structure can be identified with the action of the `zero
mode algebra' on an arbitrary highest weight state.

Zhu's algebra captures much of the structure of the (chiral) conformal
field theory: in particular it was shown by Zhu in 1996 that the 
irreducible representations of $A$ are in one-to-one
correspondence with the representations of the full vertex operator 
algebra. A conformal field theory is thus rational (in the
above physicists sense) if Zhu's algebra is finite 
dimensional.\footnote{In the mathematics literature, a vertex operator
algebra is usually called rational if in addition every positive
energy representation is completely reducible. It has been conjectured
that this is equivalent to the condition that Zhu's algebra is
semisimple.}  

In practice, the determination of Zhu's algebra is quite complicated,
and it is therefore useful to obtain more easily testable conditions
for rationality. One of these is the so-called $C_2$ condition of Zhu:
a vertex operator algebra is $C_2$-cofinite if the 
quotient space $\H_0/{\cal O}_2(\H_0)$ is finite dimensional, 
where ${\cal O}_2(\H_0)$ is spanned by the vectors of the form
\be
V_{-n-h}(\psi) \, \chi \,,\qquad n\geq 1 \,.
\ee
It is easy to show that the $C_2$-cofiniteness condition implies that
Zhu's algebra is finite dimensional. Gaberdiel and Neitzke have shown
that every $C_2$-cofinite vertex operator algebra has a simple 
spanning set; this observation can for example be used to prove that 
all the fusion rules (see below) of such a theory are finite.

\section*{Fusion rules and Verlinde's formula}

As we have explained above the correlation function of three primary
fields is determined up to an overall constant. One important question
is whether this constant actually vanishes or not since this
determines the possible `couplings' of the theory. This information is
encoded in the so-called fusion rules of the theory. More precisely
the fusion rules $N_{ij}{}^{k}\in\Nop_0$ determine the multiplicity
with which the representation of the vertex operator algebra labelled
by $k$ appears in the operator product expansion of the
two representations labelled by $i$ and $j$. 

In 1988 Verlinde found a remarkable relation between the fusion rules
of a vertex operator algebra and the modular transformation properties
of its characters. To each irreducible representation $\H_i$ of a
vertex operator algebra one can define the character
\be
\chi_i(\tau) = \hbox{Tr}_{\H_i} \left(q^{L_0 - \frac{c}{24}} \right)
\,, \qquad q= e^{2\pi i \tau} \,.
\ee
For rational vertex operator algebras (in the mathematical sense)
these characters transform under the modular transformation
$\tau\mapsto -1/\tau$ as  
\be
\chi( - 1 / \tau ) = \sum_j S_{ij} \chi_j(\tau)
\,,
\ee
where $S_{ij}$ are constant matrices. Verlinde's formula then states
that, at least for unitary theories,
\be
N_{ij}{}^{k} = \sum_l \frac{S_{il} \, S_{jl}\, S^\ast_{kl}}{S_{0l}}
\,, 
\ee
where the `0'-label denotes the vacuum representation. A general
argument for this formula has been given by Moore and Seiberg in 1989;
very recently this has been made more precise by Huang.

\section*{Modular invariance and the conformal bootstrap}

Up to now we have only considered conformal field theories on the
sphere. In order for the theory to be also well-defined on higher 
genus surfaces, it is believed that the only additional requirement
comes from the consistency of the torus amplitudes. In 
particular, the vacuum torus amplitude must only depend on the
equivalence class of tori that is described by the modular parameter
$\tau \in \mathbb{H}$, up to the discrete identifications that are
generated by the usual action of the modular group SL$(2,\Zop)$
on the upper half plane $\mathbb{H}$. For the theory with
decomposition (\ref{decomp}) this requires that the function 
\be
Z(\tau,\bar\tau) = \sum_{ij} M_{ij}\, \chi_i(\tau)\, 
\chi_j(\bar\tau)
\ee
is invariant under the action of SL$(2,\Zop)$. This is a very powerful
constraint on the multiplicity matrices $M_{ij}$ that has been
analyzed for various vertex operator algebras. For example, 
Cappelli, Itzykson and Zuber have shown that the modular
invariant WZW models corresponding to the group SU(2) have an A-D-E 
classification. The case of SU(3) was solved by Gannon, using 
the Galois symmetries of these rational conformal field theories.

The condition of modular invariance is relatively easily testable, but
it does not, by itself, guarantee that a given space of states $\bH$
comes from a consistent conformal field theory. In order to construct
a consistent conformal field theory one needs to solve the 
{\it conformal bootstrap}, that is one has to determine all the
normalisation constants of the correlators so that the resulting set
of correlators are local and factorize appropriately into 3-point 
correlators (crossing symmetry). This is typically a difficult problem
which has only been solved explicitly for rather few theories, for
example the minimal models. Recently it has been noticed that 
the conformal bootstrap can be more easily solved for the 
corresponding boundary conformal field theory. Furthermore, Fuchs,
Runkel and Schweigert have shown that any solution of the boundary
problem induces an associated solution for conformal field theory on
surfaces without boundary. This construction  relies heavily on the
relation between two-dimensional conformal field theory and
three-dimensional topological field theory (Turaev 1994).


\section*{Further Reading}
\begin{itemize}
\item[] P.~Ginsparg, {\it Applied conformal field theory}, Lectures
given at the Les Houches summer school in theoretical physics 1988. 
\item[] P.~Goddard, D.I.~Olive, {\it Kac-Moody and Virasoro algebras, 
a reprint volume for physicists}, World Scientific (1988).
\item[] P.~Goddard, {\it Meromorphic conformal field theory},
in: `Infinite dimensional Lie algebras and Lie groups: Proceedings  
of the CIRM Luminy Conference, 1988' (World Scientific, Singapore,
1989) 556.
\item[] P.~Di Francesco, P.~Mathieu, D.~S\'en\'echal,
{\it Conformal Field Theory}, Springer Verlag New York (1997).
\item[] T.~Gannon, {\it Monstrous moonshine and the classification of
CFT}, \\ arXiv:math.QA/9906167.
\item[] K.~Gawedzki, {\it Lectures on conformal field theory}, in:
Quantum fields and strings: a course for mathematicians, Vol. 2, AMS,
Providence, RI (1999).
\item[] M.R.~Gaberdiel, {\it An introduction to conformal field
theory}, Rept.\ Prog.\ Phys.\ {\bf 63} (2000) 607,
arXiv:hep-th/9910156.
\item[] C.~Schweigert, J.~Fuchs, J.~Walcher, {\it Conformal field
theory, boundary conditions and applications to string theory},
arXiv:hep-th/0011109.
\item[] A.~Pressley, G.B.~Segal, {\it Loop groups}, 
Clarendon Press Oxford (1986). 
\item[] I.~Frenkel, J.~Lepowski, A.~Meurman, 
{\it Vertex operator algebras and the Monster}, Academic Press,
Boston, MA (1988). 
\item[] V.G.~Turaev, {\it Quantum Invariants of Knots and
3-Manifolds}, de Gruy\-ter, Berlin, New York (1994). 
\item[] V.G.~Kac, {\it Vertex algebras for beginners}, AMS,
Providence, RI (1998). 
\item[] T.~Gannon, {\it Moonshine beyond the Monster: the bridge
connecting algebra, modular forms and physics}, Cambridge University
Press, to appear (2006). 
\end{itemize}

\end{document}